\documentclass{article}

\usepackage{arxiv}

\usepackage[utf8]{inputenc} % allow utf-8 input
\usepackage[T1]{fontenc}    % use 8-bit T1 fonts
\usepackage{hyperref}       % hyperlinks
\usepackage{url}            % simple URL typesetting
\usepackage{booktabs}       % professional-quality tables
\usepackage{amsfonts}       % blackboard math symbols
\usepackage{nicefrac}       % compact symbols for 1/2, etc.
\usepackage{microtype}      % microtypography
\usepackage{lipsum}
\usepackage{graphicx}
\graphicspath{ {./images/} }
\usepackage{amsmath}
\usepackage{subcaption}

\title {A novel visual data-based diagnostic approach for estimation of regime transition in pool boiling}

\author{
 Pranay Nirapure \\
  Department of Mechanical Engineering\\
  Binghamton University\\
  Binghamton, NY 13905 \\
  \texttt{pnirapu1@binghamton.edu} \\
   \And
 Ayushman Singh \\
  Department of Mechanical Engineering\\
  Binghamton University\\
  Binghamton, NY 13905 \\
  \texttt{asing133@binghamton.edu} \\
  \And
 Srikanth Rangarajan \\
  School of Systems Science and Industrial Engineering\\
  Binghamton University\\
  Binghamton, NY 13905 \\
  \texttt{srangar@binghamton.edu} \\
  \And
   Bahgat Sammakia \\
  Department of Mechanical Engineering\\
  Binghamton University\\
  Binghamton, NY 13905 \\
  \texttt{bahgat@binghamton.edu} \\
}

\begin{document}
\maketitle
\begin{abstract}
This study introduces a novel metric, the Index of Visual Similarity (IVS), to qualitatively characterize boiling heat transfer regimes using only visual data. The IVS is constructed by combining morphological similarity, through SIFT-based feature matching, with physical similarity, via vapor area estimation using Mask R-CNN. High-speed images of pool boiling on two distinct surfaces, polished copper and porous foam, are employed to demonstrate the generalizability of the approach. IVS captures critical changes in bubble shape, size, and distribution that correspond to transitions in heat transfer mechanisms. The metric is validated against an equivalent metric, $\phi$, derived from measured heat transfer coefficients (HTC), showing strong correlation and reliability in detecting boiling regime transitions, including the onset of nucleate boiling and proximity to critical heat flux (CHF). Given experimental limitations in precisely measuring changes in HTC, the sensitivity of IVS to surface superheat is also examined. IVS thus emerges as a powerful, rapid, and non-intrusive tool for real-time, image-based boiling diagnostics, with promising applications in phase change heat transfer.

\end{abstract}

% \includegraphics{figs/grabs.pdf}


% Keywords

\maketitle

\section{Introduction}

Boiling heat transfer stands out as a highly efficient and appealing solution for thermal management across a wide range of applications \cite{b1}-\cite{b2}. These include critical operations such as thermal management of high-power-density electronics in data centers \cite{b3}-\cite{b4}. However, its effectiveness is fundamentally limited by a phenomenon known as boiling crisis which is marked by critical heat flux (CHF). Beyond CHF, heat transfer plummets abruptly due to the formation of a vapor blanket over the heated surface, which can lead to thermal runaway and catastrophic system failure. Therefore, for safe and efficient utilization, it is essential not only to delay the onset of CHF but also to accurately estimate heat flux well before reaching this limit \cite{b5}-\cite{b7}.

\begin{figure}[htbp]
    \centerline{\includegraphics[scale=0.5]{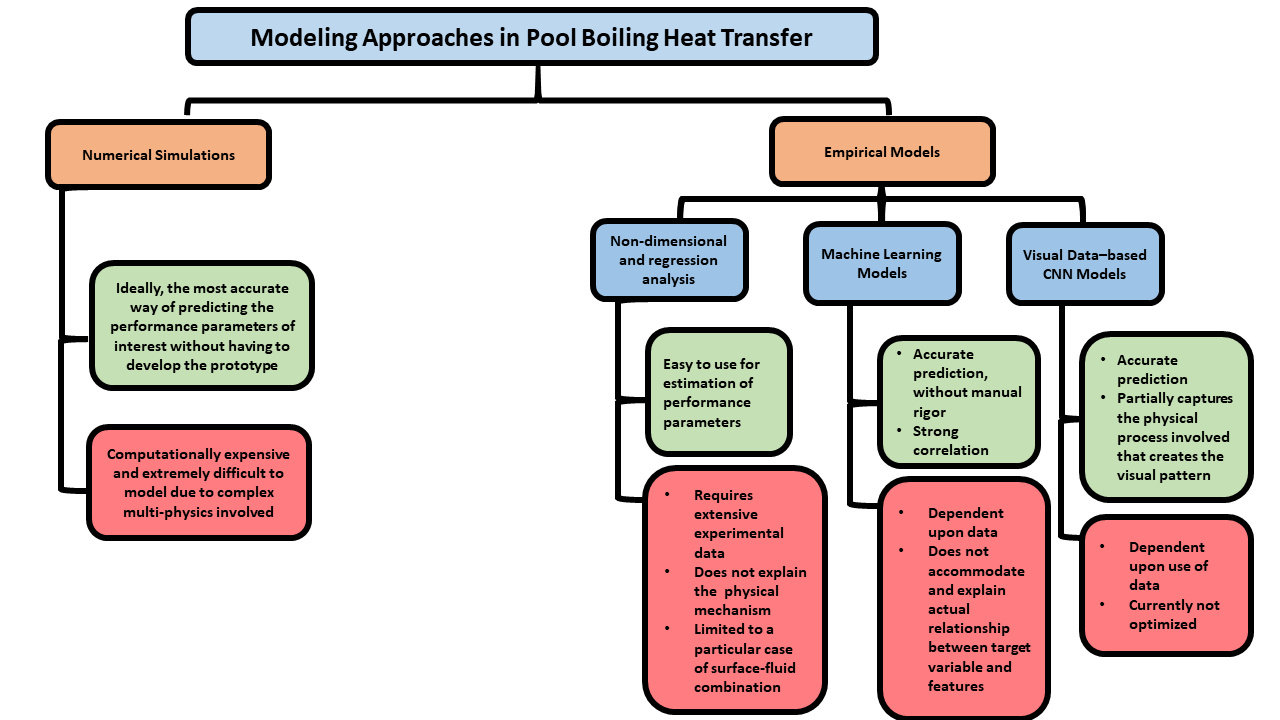}}
    \caption{Classification of boiling heat transfer modeling approaches, highlighting the evolution toward data-centric diagnostics.}
    \label{fig1}
\end{figure}

Achieving this, however, remains a major challenge due to the inherently complex and multi-scale nature of boiling. It involves the simultaneous interaction of various physical phenomena across different length and time scales, making it extremely difficult to isolate and understand the underlying heat transfer mechanisms\cite{b8}. To date, there is no consensus within the boiling research community regarding the dominant mechanisms governing heat transfer in boiling processes \cite{b9}. Additionally, computational models are often inadequate due to the disparity between the complexity of the physical processes and current computational capabilities \cite{b10}. Fig. \ref{fig1} shows a comprehensive classification of various modeling approaches used to model pool boiling heat transfer.

Over the years, numerous empirical models have been developed to estimate boiling heat flux and regime based on measurable parameters such as surface temperature in addition to thermophysical properties of working fluids, and surface characteristics \cite{b11}-\cite{b18}. More recently, advanced data-driven approaches have demonstrated exceptional potential \cite{b19}-\cite{b26}. In particular, visual pattern recognition using neural networks has shown remarkable accuracy in estimating boiling heat flux, with some studies reporting prediction errors less than 10\text{\%}\cite{b27}. However, a major limitation of both empirical and machine learning-based models lies in their lack of generalizability. These models are inherently dependent on the experimental datasets used for their development, making them highly specific to particular fluid-surface combinations. As a result, their predictive capability is constrained to those specific conditions and cannot be reliably extended to new scenarios without retraining or recalibration. 

This challenge is rooted in the fundamental role that fluid-surface interaction plays in boiling heat transfer. Variations in surface characteristics (such as roughness, wettability, and material composition) as well as differences in fluid properties, significantly influence both the visual boiling patterns and the dominant heat transfer mechanisms. Consequently, changes in the fluid or surface can lead to deviations in visual features and the way they are used in existing models, limiting their generalizability \cite{b28}-\cite{b29}. A general predictive model should seek common characteristics that quantify change in boiling regime or heat flux regardless of the surface-fluid combination.
Although several efforts have been made to develop more accurate models for estimating heat flux, significant progress toward a general model applicable across surface-fluid combinations remains limited.

To address these challenges, this work lays the foundation for a novel approach to estimating changes in the boiling regime. Rather than directly correlating visual patterns with heat flux values, we propose a relative analysis of these patterns across various boiling conditions. This methodology enables the development of a framework that asses the visual signature of boiling regime change for two tested surface-fluid combinations. Central to this approach is the introduction of a parameter termed the Index of Visual Similarity (IVS), which serves as a comparative metric for assessing changes in boiling behavior visually. The interpretation and application of IVS are demonstrated as part of a future generalized framework for monitoring and estimating heat transfer across the full spectrum of pool boiling, from the onset of nucleate boiling to the critical heat flux condition.

%\section{Problem formulation}

\section{Results} \label{res}

Visual data-based empirical models have shown superior performance owing to their ability to extract features that implicitly represent the complex, multi-physics nature of boiling heat transfer through distinct visual patterns. These models incorporate the influence of phenomena such as chaotic nucleation, intricate bubble interactions, and phase change behavior. The visual patterns observed during boiling are a direct manifestation of these underlying processes, making them a rich source of information for quantifying heat flux. As such, image-based approaches offer a unique advantage by leveraging the signature of energy transport mechanisms embedded within the visual data.

\begin{figure}[htbp]
    \centerline{\includegraphics[scale=0.5]{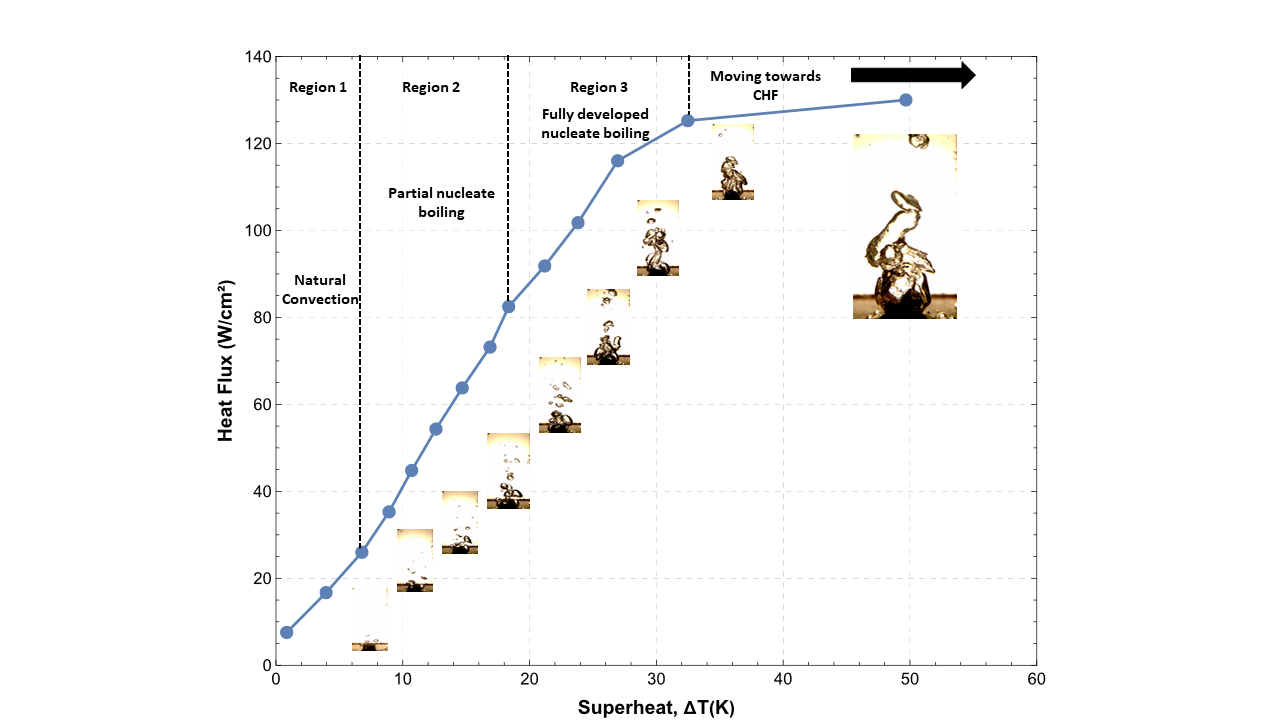}}
    \caption{Pool boiling curve: simultaneous visualization of bubble dynamics with increasing heat flux on a porous Cu foam surface, depicting distinct transitions in boiling regimes}
    \label{fig2}
\end{figure}

The relationship between the heat transfer coefficient and heat flux is not straightforward due to the simultaneous contribution of multiple mechanisms. As highlighted in numerous studies, each of the five primary mechanisms responsible for transferring heat from the surface into the fluid (namely, liquid convection, vapor convection, resetting, contact line evaporation, and micro-layer evaporation) contributes differently with change in heat flux  \cite{b30,b9}. The relative contribution of each mechanism varies with changes in heat flux, directly influencing the generation of vapor and, consequently, the visual characteristics of the vapor. This forms the foundational principle for the approach employed in this work. As shown in fig. \ref{fig2}, for porous foam surface, with a change in heat flux, visual pattern changes. The figure presents a qualitative classification of different stages of boiling, identified through systematic observation of slope transitions in the boiling curve.

\subsection{Index of visual similarity}

The Index of Visual Similarity (IVS) is introduced to quantify the change in visual patterns across varying heat fluxes by measuring the degree of similarity between a pair of heat fluxes. It serves as a tool to track the rate of change in specific visual features as heat flux changes. IVS is computed for a pair of heat fluxes, $Q_{n}$ and $Q_{n+1}$,which are separated by $\Delta Q$. The IVS is then defined at the midpoint,Q= ($Q_{n}$+$Q_{n+1}$)/2. A detailed procedure for calculating the IVS can be found in section \ref{IVS}.

These visual features are extracted using two distinct reasoning approaches, namely morphological and physical properties of bubbles, which together constitute the subcomponents of the IVS:

\subsubsection{Morphological similarity}
\begin{figure}[htbp]
    \centerline{\includegraphics[scale=0.5]{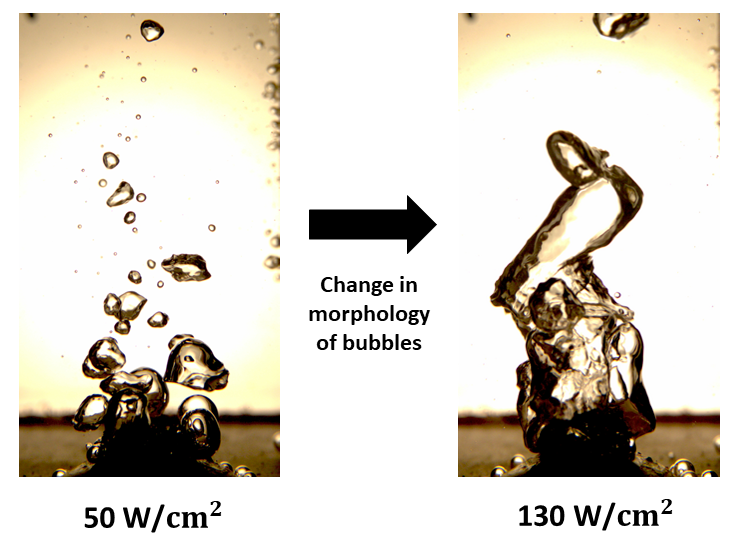}}
    \caption{Morphological evolution of vapor bubbles under increasing heat flux for Dataset-2, illustrating transitions from spherical to elongated columnar structures due to lateral coalescence}
    \label{fig3}
\end{figure}

Morphological similarity measures the extent of common features between images corresponding to heat fluxes to be compared. These features represent the 2D morphological properties of bubbles, including shape, form, and proximity with other bubbles. Higher commonality between features represents a higher value of morphological similarity. It is used to extract features that are harder to quantify as they do not have a quantifiable marker. From a physical standpoint, as illustrated in Fig.\ref{fig3}, bubble morphology undergoes a significant transition with increasing heat flux. At lower heat fluxes, bubbles tend to retain a nearly spherical geometry due to a balance between surface tension force and vapor momentum force. As heat flux approaches the critical heat flux (CHF) threshold, vapor momentum increasingly overcomes surface tension, resulting in lateral coalescence and the emergence of elongated, columnar bubble structures \cite{b34}.

These visually identifiable morphological properties, though challenging to track manually and often lacking distinct markers, play a critical role in characterizing essential physics. Consequently, this similarity metric captures all morphological changes in bubble appearance, providing a way to gauge changes in the stages of boiling.

\subsubsection{Physical similarity}
The rationale for incorporating this similarity lies in its direct connection to the fundamental understanding of phase change processes. It quantifies changes in the detected bubble area by comparing two heat fluxes. In a 2D image, the bubble area serves as a rough approximation of vapor volume. As heat transfer from the surface to the fluid is influenced by contact line evaporation, the vapor volume changes accordingly, leading to alterations in bubble area \cite{b35}. These changes, though subtle, are not easily detectable through manual observation. While morphological similarity captures visual features that may be challenging to interpret intuitively, physical similarity offers a reality check, providing a more direct connection between visual features and the extent of phase change heat transfer. However, physical similarity is solely dependent on changes in vapor volume and does not account for morphological properties, which are essential for representing other physical changes. Therefore, both morphological and physical similarities are equally important and must be considered together to accurately represent the visual information.

\subsection{Results from IVS}
\begin{figure}[htbp]
    \centerline{\includegraphics[scale=0.26]{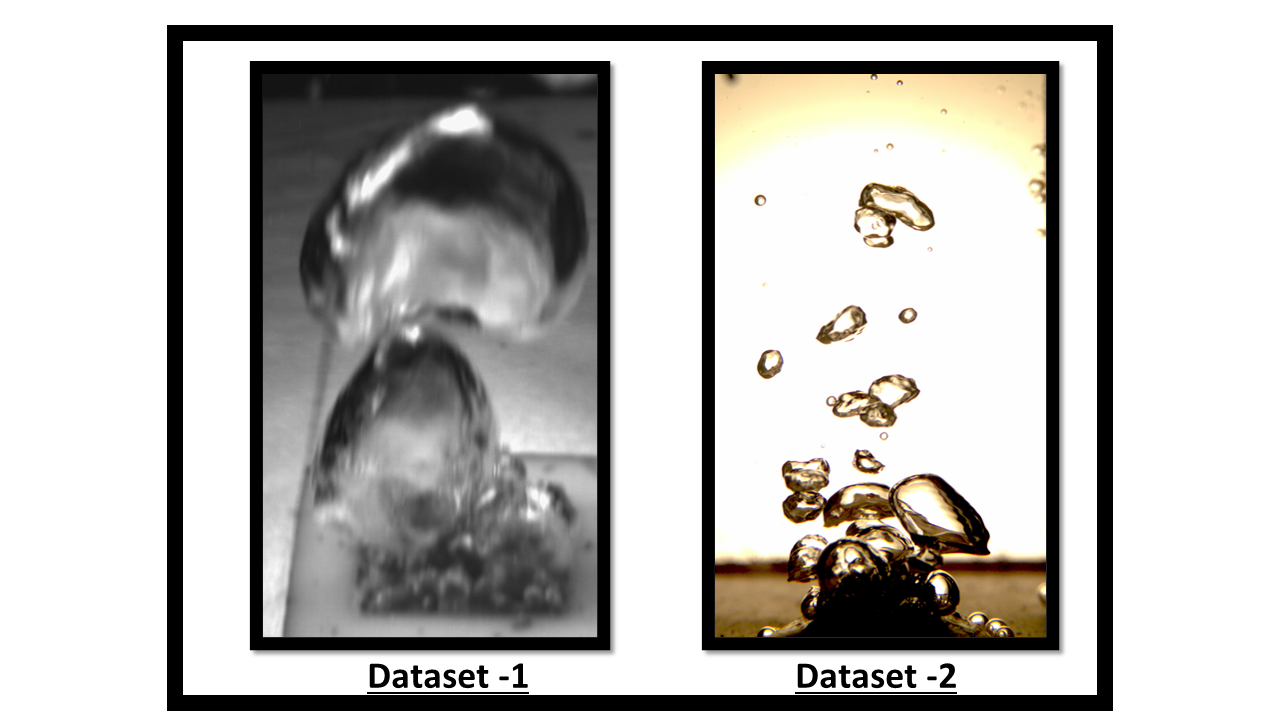}}
    \caption{Representative sample images from Dataset-1 and Dataset-2, captured at 60 $W/cm^2$, highlighting surface differences (polished copper vs. porous Cu foam) and visualization techniques}
    \label{fig4}
\end{figure}

Two distinct surfaces, each employing entirely different visualization techniques, have been utilized to establish a general approach that is independent of surface-fluid interactions. The rationale behind this is to ensure that any observed trends are not coincidental or biased toward a particular surface or visualization method, but rather reflect an actual change in visual features. Representative sample images are shown in fig. \ref{fig4}. For detailed information on the surface properties and visualization techniques, please refer to section \ref{methods}.

\begin{figure}[htbp]
    \centerline{\includegraphics[scale=0.30]{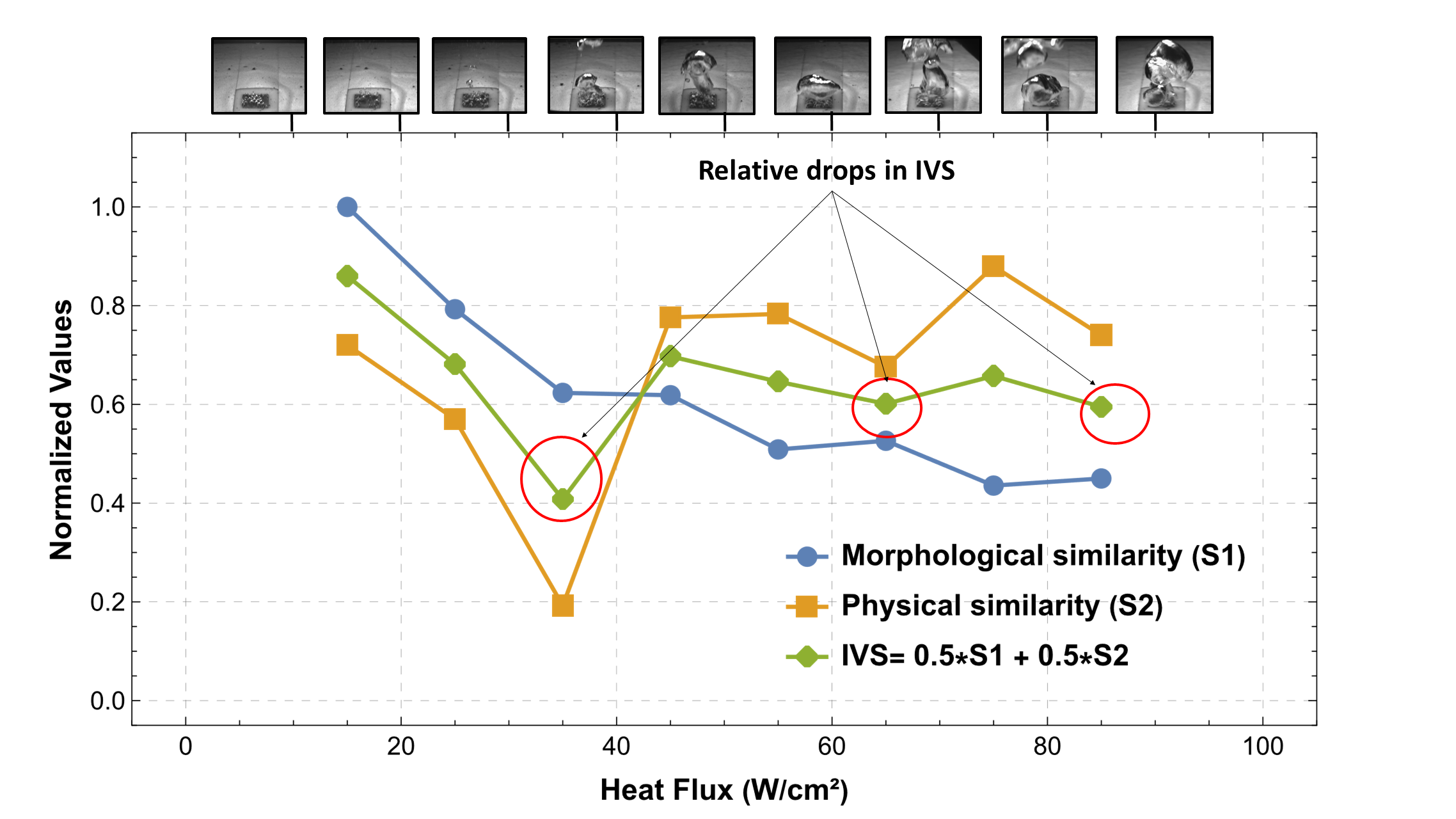}}
    \caption{Variation of morphological similarity, physical similarity, and the resulting Index of Visual Similarity (IVS) for Dataset-1, accompanied by representative bubble evolution patterns. Notable drops in IVS correspond to significant transitions in visual features, aligning with regime changes that are readily identifiable through manual inspection of the accompanying images.}
    \label{fig5}
\end{figure}

Fig. \ref{fig5} illustrates the variation in the IVS for Dataset-1, accompanied by a sample image (randomly selected from the captured image dataset) to represent the corresponding visual pattern.
\begin{figure}[htbp]
    \centerline{\includegraphics[scale=0.30]{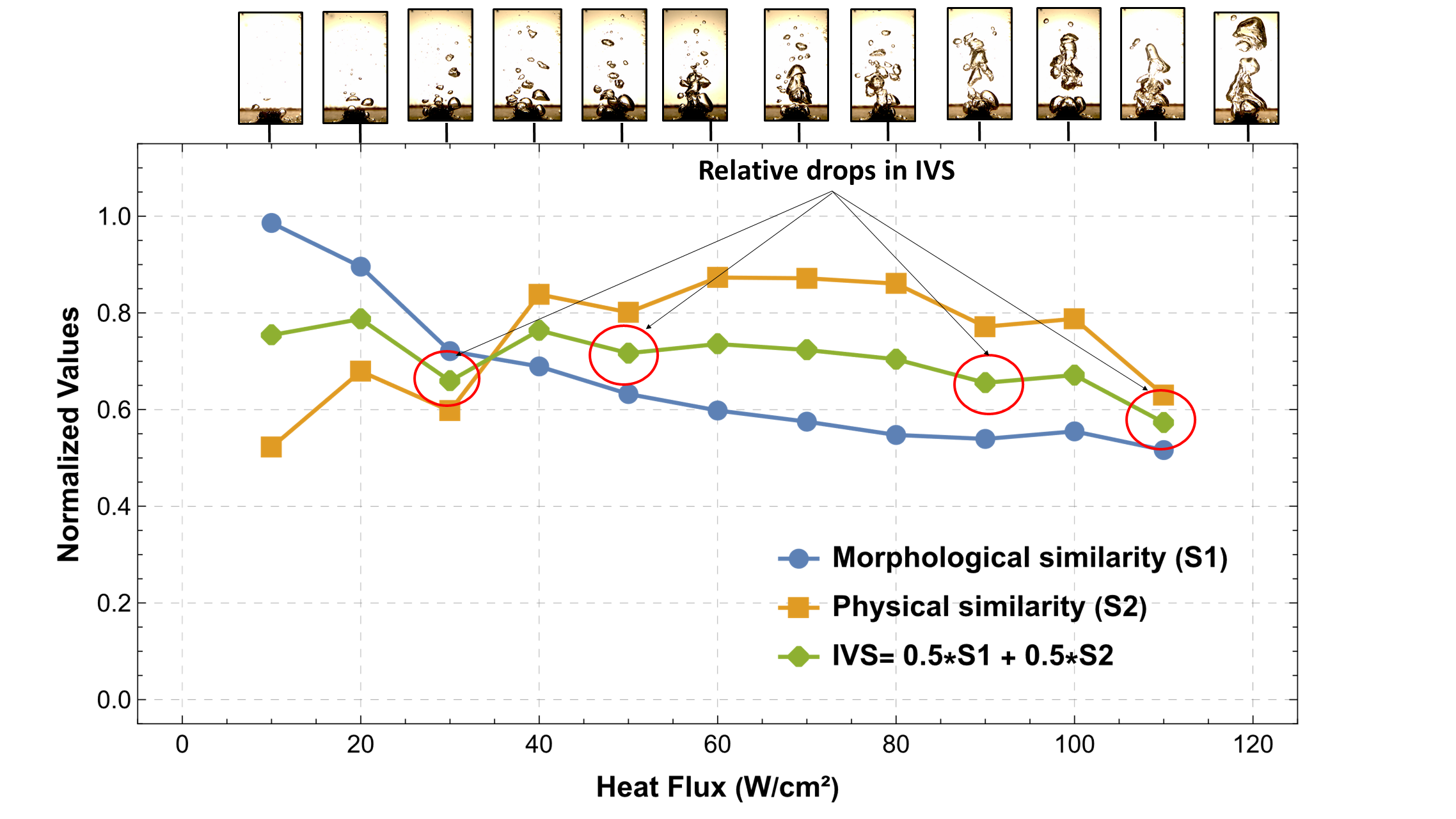}}
    \caption{Variation of morphological similarity, physical similarity, and the resulting Index of Visual Similarity (IVS) for Dataset-2, demonstrating the consistency and robustness of the framework across different surface types. The observed trends reaffirm the framework's applicability in capturing regime transitions independent of surface morphology.}
    \label{fig6}
\end{figure}

It can be observed that the trend in the IVS aligns well with noticeable changes in visual patterns, particularly in regions where these changes are easily discernible through manual inspection. In Dataset 1, for example, as the heat flux increases from 10 W/cm² to 30 W/cm², there is a marked increase in bubble nucleation, which corresponds to a distinct drop in IVS. In this range, morphological and physical similarity exhibit similar trends. However, certain deviations highlight their complementary nature. Notably, between 70 W/cm² and 80 W/cm², physical similarity remains relatively high, while morphological similarity drops significantly. This discrepancy illustrates the enhanced sensitivity of morphological similarity in capturing subtle changes in bubble shape and form—features that may evolve even when the overall vapor area remains nearly constant, thus yielding high physical similarity.

A similar observation holds for Dataset 2 as shown in fig. \ref{fig6}, further reinforcing the robustness of the proposed method. It is important to emphasize that identical trends between morphological and physical similarity are not necessarily expected, given that each captures distinct aspects of the visual information. Therefore, despite occasional divergence in their trends, incorporating both metrics is essential for a comprehensive and reliable representation of boiling physics.

\subsection{Physical interpretation}

Until this point, IVS may appear abstract, lacking direct physical validation to confirm its relevance to heat transfer characteristics. This subsection addresses that gap by establishing a qualitative correlation between IVS and the heat transfer coefficient (HTC), as measured simultaneously during image acquisition. IVS can be interpreted as a metric reflecting changes in visual patterns. Fundamentally, these changes arise from variations in bubble statics (shape, size, and form), which is a direct manifestation of the underlying heat transfer processes governed by HTC and heat flux. Therefore, any significant change in heat transfer should correspond to a detectable shift in the visual pattern.

To substantiate this relationship, thermocouple data are used to compute a metric $\phi$ : the complement of relative change in the HTC across increasing heat flux. This metric captures the sensitivity of boiling HTC to heat flux: a lower value indicates more rapid changes in HTC over small increments in heat flux. A detailed explanation of the calculation methodology and the associated statistical analysis is provided in section \ref{phi}. This comparative analysis reinforces the physical basis of IVS, demonstrating its utility as a proxy for identifying transitions in heat transfer behavior through visual observations.

\begin{figure}[htbp]
    \centerline{\includegraphics[scale=0.40]{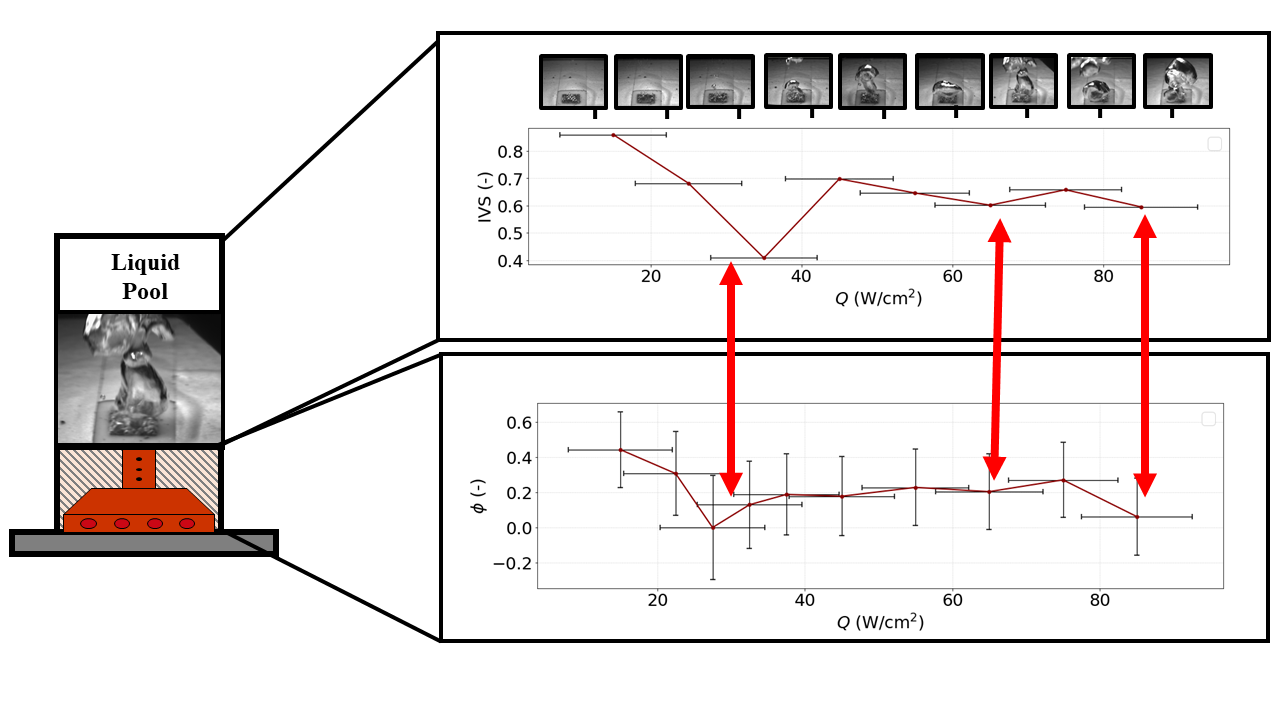}}
    \caption{Comparison of IVS with complement of change in heat transfer coefficient (HTC) for Dataset-1, showing strong correspondence with regime transitions in pool boiling}
    \label{fig7}
\end{figure}
\begin{figure}[htbp]
    \centerline{\includegraphics[scale=0.40]{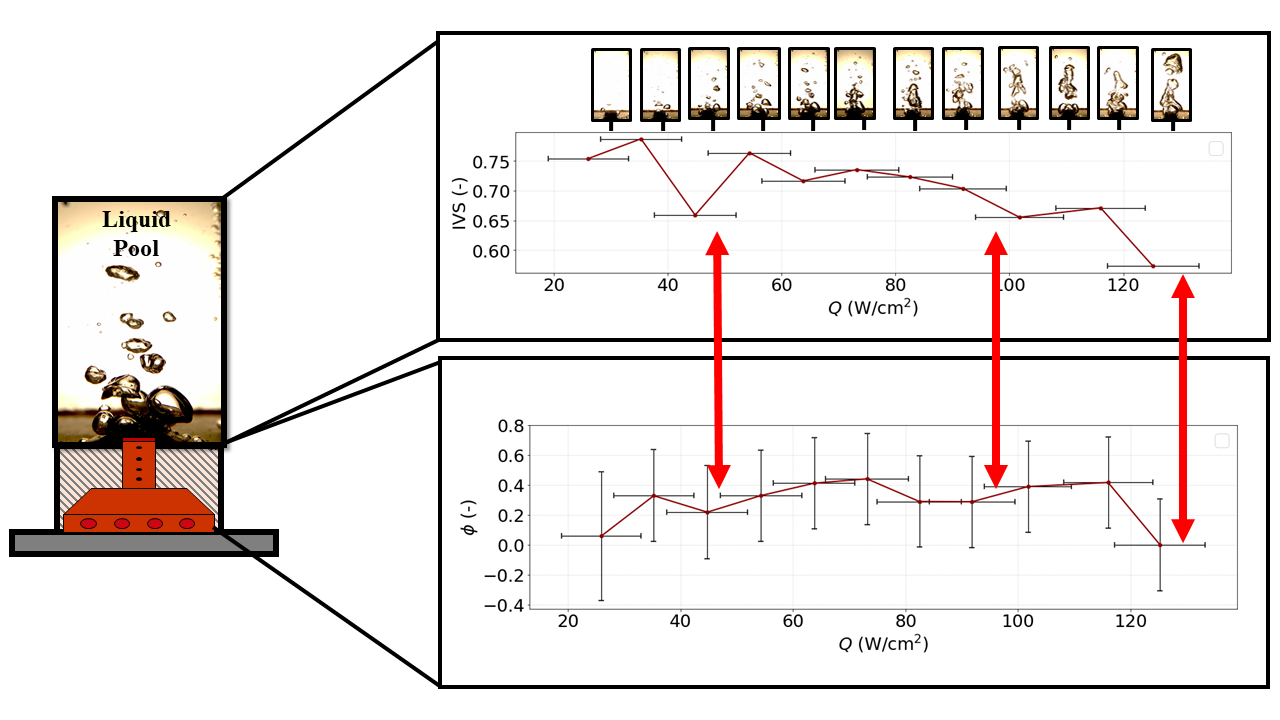}}
    \caption{Comparison of IVS and complement of change in heat transfer coefficient for Dataset-2, reaffirming the correlation between visual similarity and thermal behavior across surface-fluid conditions}
    \label{fig8}
\end{figure}

Fig. \ref{fig7} presents a comparison between the IVS and the complement of relative change in HTC, revealing strong agreement at all major transitions in the visual boiling pattern. The first significant shift occurs near 30 W/cm², signaling the end of the natural convection-dominated regime (Region 1). Beyond this point, the boiling process transitions into the partial nucleate boiling regime (Region 2). A subsequent drop in both IVS and the inverse HTC fluctuation around 65 W/cm² marks the onset of fully developed nucleate boiling (Region 3). Toward the end of this regime, another notable decline is observed in both metrics, indicating proximity to the CHF. This final drop is primarily attributed to a relative shift in the contributions of contact line evaporation and vapor convection to the overall heat transfer mechanism.

These observations reinforce the physical validity of IVS, demonstrating its ability to reliably capture variations in the HTC based solely on visual information. Given the challenges associated with directly measuring HTC (particularly the need for precise, localized surface temperature data) IVS emerges as a compelling, non-intrusive alternative for quantifying and monitoring boiling performance.

Crucially, the alignment between IVS and HTC fluctuation is not limited to a single surface-fluid combination. As shown in fig. \ref{fig8}, similar trends are observed across two distinct surface types and visualization techniques, underscoring the robustness and generalizability of this approach.

An important consideration in this analysis is the unavoidable uncertainty inherent in temperature measurement and the subsequent uncertainty in $\phi$. In the present experimental setup, HTC is estimated using a copper block instrumented with three thermocouples to approximate the surface temperature gradient. This introduces a significant error in the estimation of the derived quantity $\phi$. In fact, the propagated uncertainty often matches or exceeds the magnitude of the observed change in HTC. Consequently, the limitations of the instrumentation is a fundamental barrier to detecting fine-grained transitions in boiling . Under such conditions, isolating genuine change in boiling regime becomes exceedingly difficult using HTC alone unless highly accurate instrumentation is used.

This makes the observed alignment between IVS and the derived $\phi$ all the more compelling. Despite the limitations of direct measurement, IVS consistently captures transitions in heat transfer regimes through purely visual cues. Its robustness against measurement noise highlights its potential as a non-intrusive, data-driven proxy for detecting changes in boiling dynamics. Without substantial upgrades, such as the use of thin-film heaters and higher-resolution infrared thermography, these uncertainties in $\phi$ remain intrinsic to the methodology. Hence, IVS stands out as a uniquely valuable tool for thermal regime identification, offering reliable solution where conventional measurements fall short.

\begin{figure}[htbp]
\centerline{\includegraphics[scale=0.30]{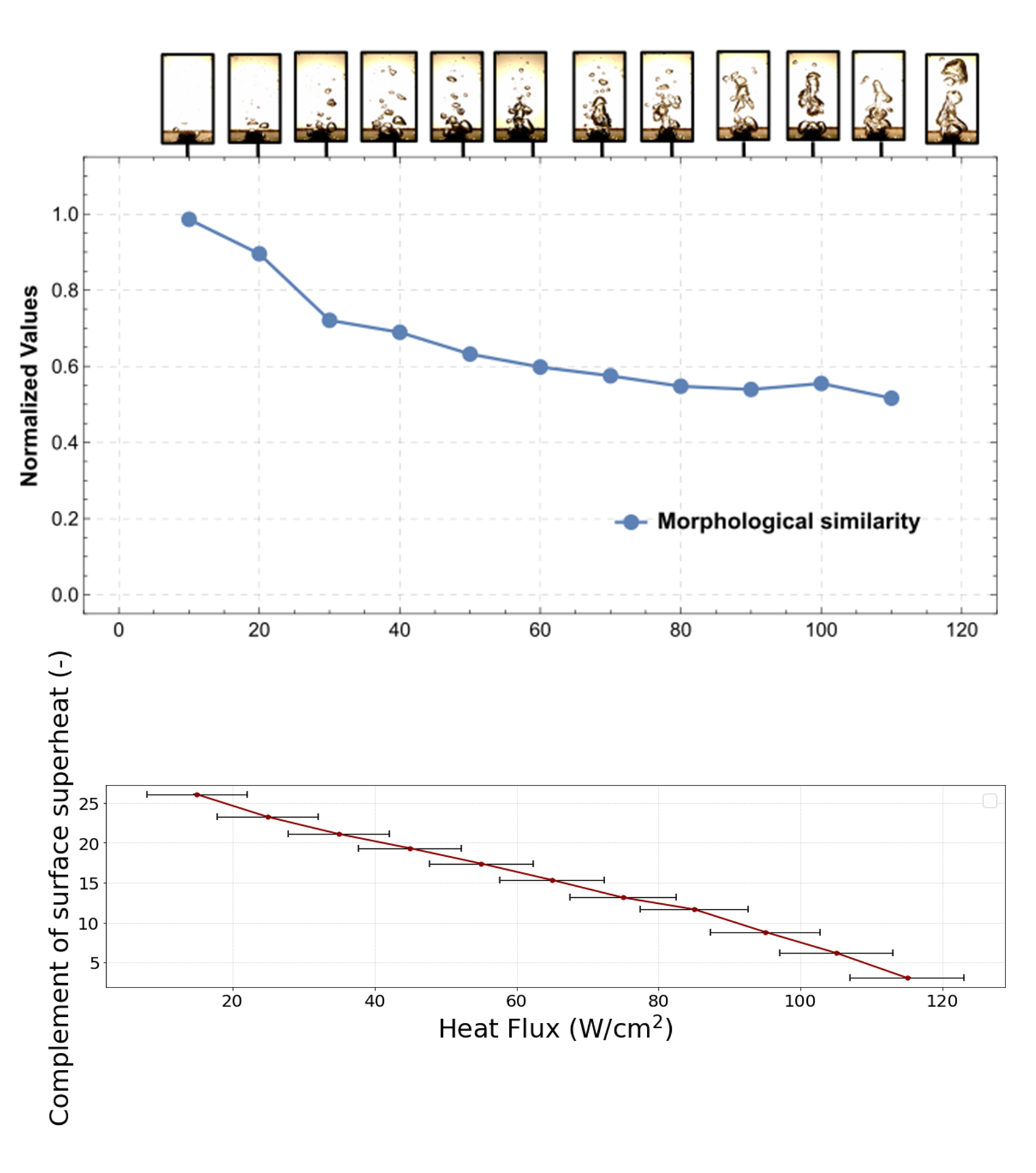}}
\caption{Comparison of morphological similarity and the complement of surface superheat for Dataset-2}
\label{fig8.1}
\end{figure}

Another noteworthy observation is the ability of morphological similarity to capture the influence of surface superheat variations, as illustrated in Fig. \ref{fig8.1}.
This alignment underscores the potential of IVS to encode meaningful physical transitions within the boiling process, particularly those driven by changes in surface thermal conditions. While these results are promising, comprehensive validation using advanced diagnostics, such as high-resolution infrared thermography and thin-film heaters remains essential to fully establish the robustness of IVS.

\section{Discussions}
The ability of the IVS to detect subtle changes in visual patterns, along with its strong agreement with variations in the HTC and surface superheat, establishes it as a powerful tool for identifying and correlating visual features with distinct boiling regimes. It is well established that abrupt changes in HTC are indicative of regime transitions in boiling heat transfer \cite{b31}. Consequently, sharp variations in IVS can be directly interpreted as transitions between boiling regimes.

As outlined in section \ref{res}, several mechanistic models have been proposed in the past to describe the principal modes of energy transport from the heated surface to the bulk fluid. Prior studies have examined the relative contribution of each mechanism across different boiling regimes. At lower heat fluxes and under subcooled or saturated conditions, single-phase convection predominates. As the system enters the nucleate boiling regime, energy transport becomes increasingly governed by contact line evaporation \cite{b9}. These shifts in dominant mechanisms influence both the morphology and dynamics of vapor generation, which in turn are encoded in the visual patterns captured by IVS. These changes, which are traditionally difficult to detect (without intrusive measurements or the use of a simultaneous phase and temperature map), are effectively quantified through IVS.

In this context, the present work lays the foundation for a novel, vision-based framework capable of identifying crucial events in boiling through non-intrusive visual analysis. By linking visual features directly to regime transitions and heat transfer characteristics, IVS offers a promising solution toward generalized, data-driven diagnostics in phase change heat transfer.

\section{Current limitations and future work}
This study does not attempt to directly estimate the boiling heat flux; rather, it focuses on capturing transitions between boiling regimes. The ability to identify these regime shifts is a critical advancement, particularly for applications involving real-time monitoring and early prediction of CHF. By successfully quantifying visual transitions linked to underlying change in physical quantities like HTC and surface superheat, this approach provides a strong foundation for future developments. Moving forward, we aim to expand upon this framework by further interpreting the IVS with the goal of enabling direct, quantitative estimation of heat flux. Concurrently, we plan to integrate advanced diagnostic tools such as high-resolution infrared thermography to achieve localized, low-uncertainty HTC measurements to reinforce the proposed significance of using visual features. This progression has the potential to yield a generalized, non-intrusive methodology for real-time diagnostics and control in boiling systems.

\section{Methods} \label{methods}
\subsection{Experimental Setup}

\begin{figure*}[htbp]
    \centerline{\includegraphics[scale=0.5]{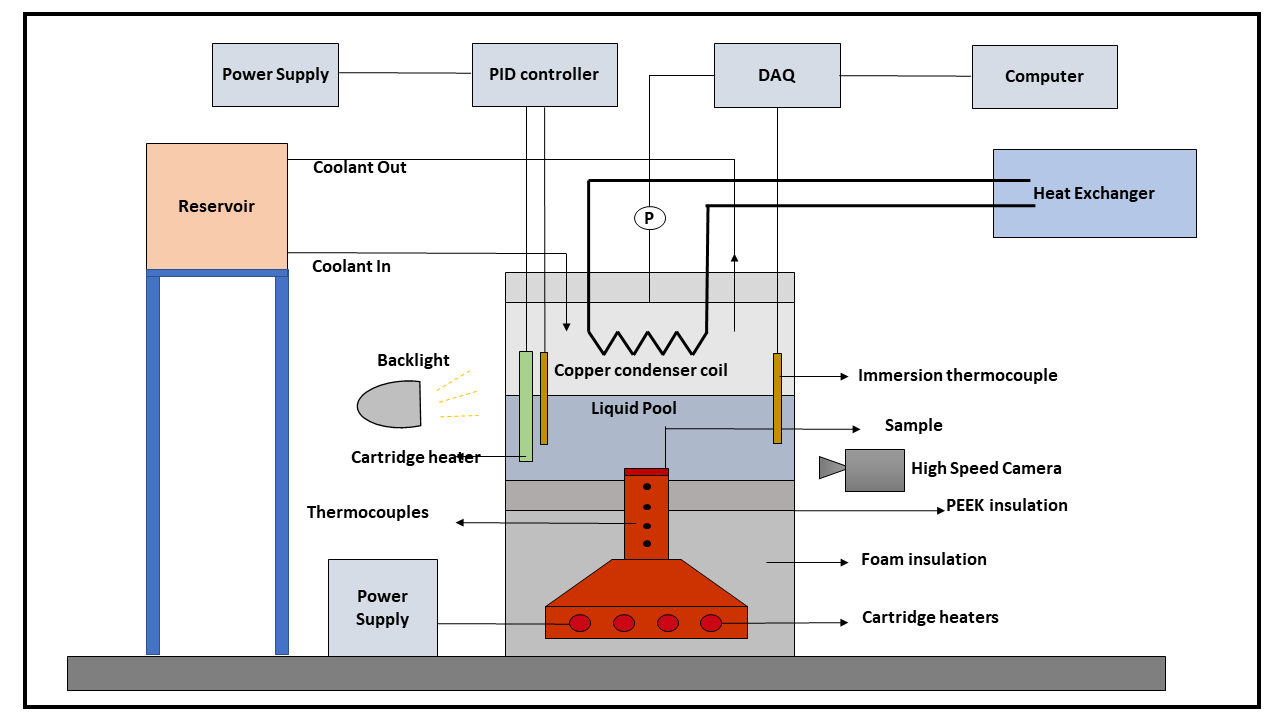}}
    \caption{Schematic of the experimental setup used for visual data acquisition and pool boiling measurements}
    \label{fig9}
\end{figure*}

Fig. \ref{fig9} overviews the experimental setup used in this study. The setup consists of a copper block containing four cartridge heaters, which supply heat to a compact die area measuring 1 cm². This block is situated in a tank, and its bottom section features four polycarbonate windows (4" X 3"), providing visual access to the surface of interest and enabling backlighting for shadow-based imaging in the case of dataset-2. 

To investigate saturated pool boiling, the liquid pool is meticulously maintained at the saturation conditions (100°C at 1 atm for water) by a PID controller-based feedback loop, which uses two thermocouples immersed in the pool to continuously monitor the temperature and actuate heaters immersed in the pool accordingly.
The experimental procedure begins by heating the pool until it reaches a stable saturation temperature, typically taking about one hour to ensure uniform temperature distribution throughout the pool. Following this initial phase, the liquid is vigorously boiled for 30 minutes to remove non-condensable gases. 
Heat is then applied to the copper block starting from 5 W/cm², increasing by 2 W/cm² during the initial bubble nucleation and by 10 W/cm² after reaching 10 W/cm². Vapor generated from boiling condenses back into the pool at the surface of the copper condenser coil to maintain pressure. Cold water is continuously circulated from the heat exchanger through the condenser coil to extract waste heat. Temperature data is systematically recorded at steady state for each heat flux value with simultaneous high-speed visual data acquisition.

Dataset-1 was captured using deionized water boiling over a polished copper surface, while Dataset-2 utilized deionized water with a 3 mm thick porous foam surface containing 60 pores per inch (PPI).

\subsection{ Heat flux measurement}

To derive the actual heat flux out of the exposed surface, Fourier's law has been used and temperature has been measured at 3 different points along the length of the copper block.
\begin{figure}[htbp]
    \centerline{\includegraphics[scale=0.28]{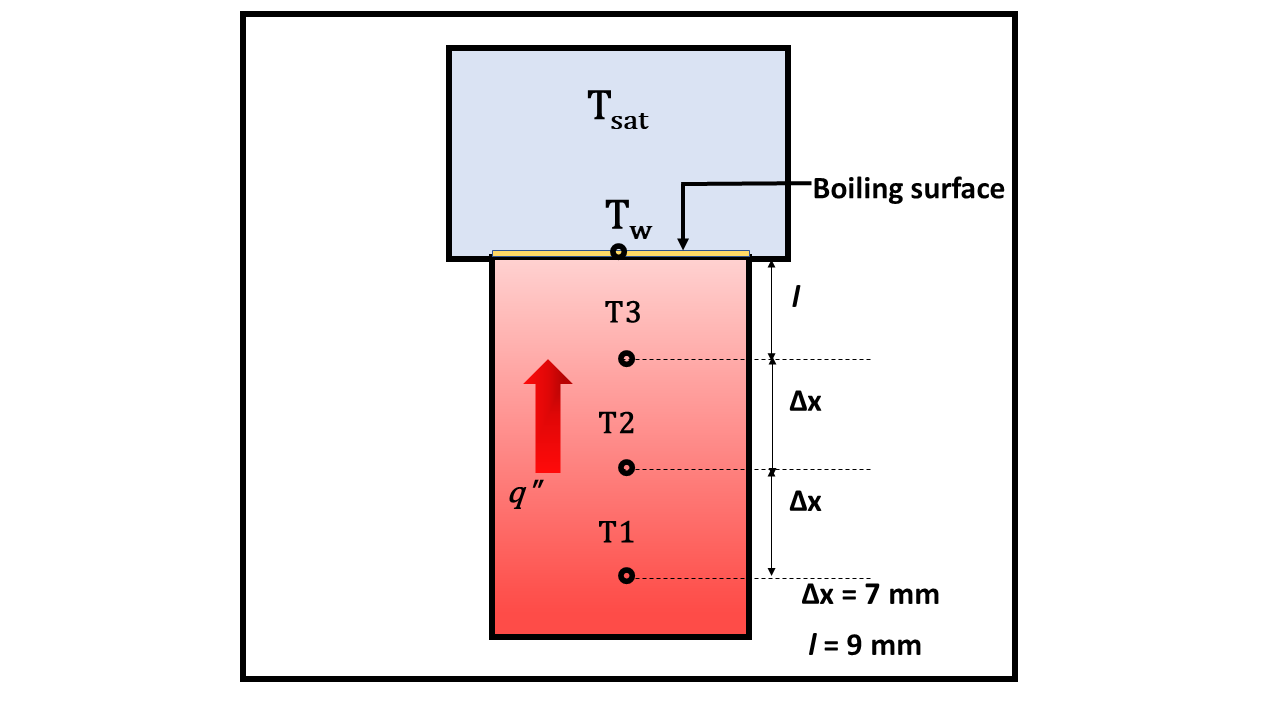}}
    \caption{Data reduction methodology used to estimate heat flux and surface temperature. The figure illustrates the placement of three thermocouples ($T_1$, $T_2$, and $T_3$) embedded along the length of the copper heating block. Using these measurements, the temperature gradient is resolved via a second-order finite difference scheme to compute the heat flux through Fourier’s law. The surface (wall) temperature, $T_w$, is then back-calculated by accounting for conduction losses through the block. The resulting wall superheat, $\Delta T_{wall} = T_w - T_{sat}$, enables computation of the heat transfer coefficient ($h$), which serves as a key metric for boiling performance evaluation.}
    \label{figad2}
\end{figure}
Fig. \ref{figad2} shows the locations of the thermocouples $T_1$, $T_2$ and $T_3$ for the temperature measurement. $T_w$ represents Wall temperature and $T_{sat}$ represents saturation temperature of liquid.
Heat flux is calculated as shown below:
$$ q^"=-k_{cu} \frac{\Delta T^*}{2\Delta x} $$
$$\Delta T^* = 3T_3-4T_2+T_1$$
$$T_{w}= T_3 - q^" . \frac{l}{k_{cu}} $$
$$\Delta T_{wall}= T_{w}- T_{sat} $$
$$ h= \frac{q^"}{\Delta T_{wall}}$$

where, $\Delta T_{wall}$: Wall Superheat, 
h: Heat transfer coefficient$(W/cm^2K)$, 
$q^"$: Input heat flux $(W/cm^2)$ and $k_{cu}$: thermal conductivity of copper.

\subsection{Visual data acquisition}
\begin{figure}[htbp]
    \centerline{\includegraphics[scale=0.25]{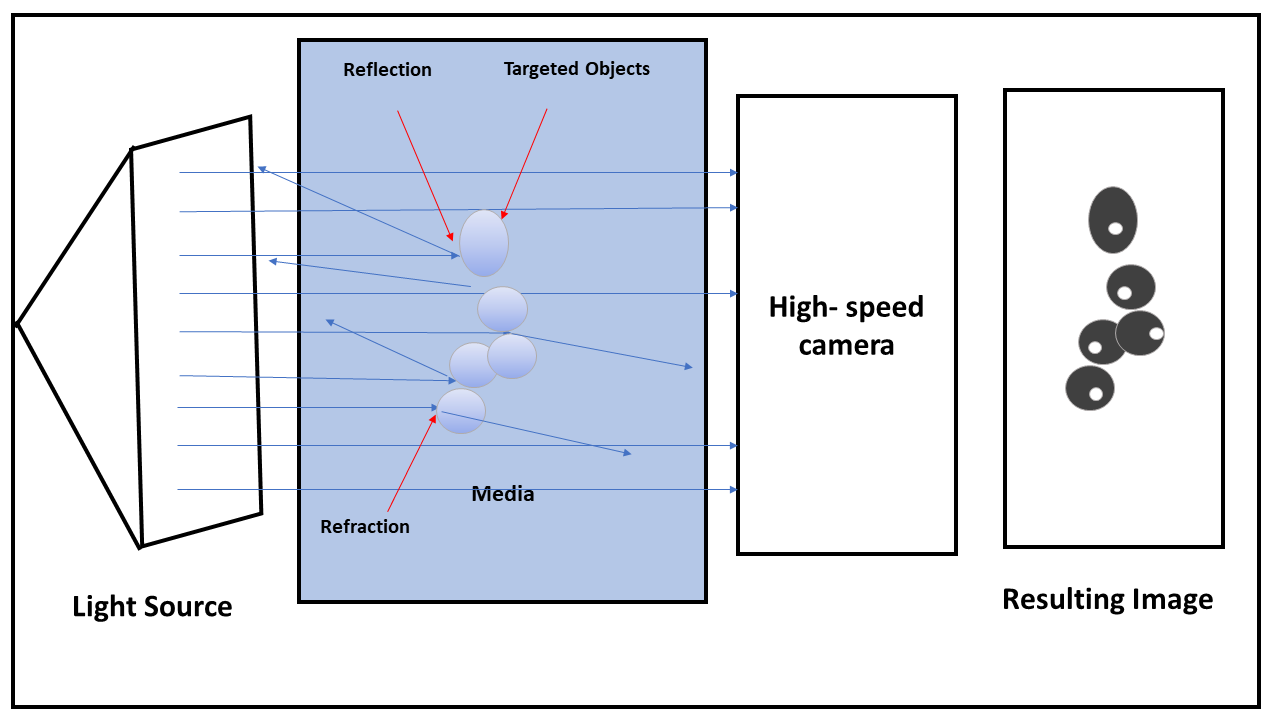}}
    \caption{Visual data acquisition setup for Dataset-2, employing shadowgraph-based high-speed imaging to enhance bubble boundary visibility}
    \label{fig10}
\end{figure}

To acquire Dataset-1, a Fastec IL-5 high-speed monochromatic camera was employed at a frame rate of 500 fps, using front-plane illumination of the heater surface. In contrast, Dataset-2 was captured using a Fastec HS-7 high-speed camera operating at 1000 fps, utilizing a shadowgraph-based visualization technique to highlight bubbles as shown in fig. \ref{fig10}.

\subsection{Index of visual similarity}\label{IVS}
IVS quantifies the degree of similarity between boiling visual patterns corresponding to two distinct heat flux values, thereby capturing the visual uncertainty associated with differentiating between them. For each heat flux condition, a series of images is recorded under steady-state conditions. IVS is defined at a representative heat flux Q, computed from a pair of heat fluxes $Q_{n}$ and $Q_{n+1}$, where \begin{equation} Q=(Q_{n}+Q_{n+1})/2 \end{equation}

\begin{equation} \Delta Q=(Q_{n+1}-Q_{n}) \end{equation}
\begin{figure}[htbp]
    \centerline{\includegraphics[scale=0.25]{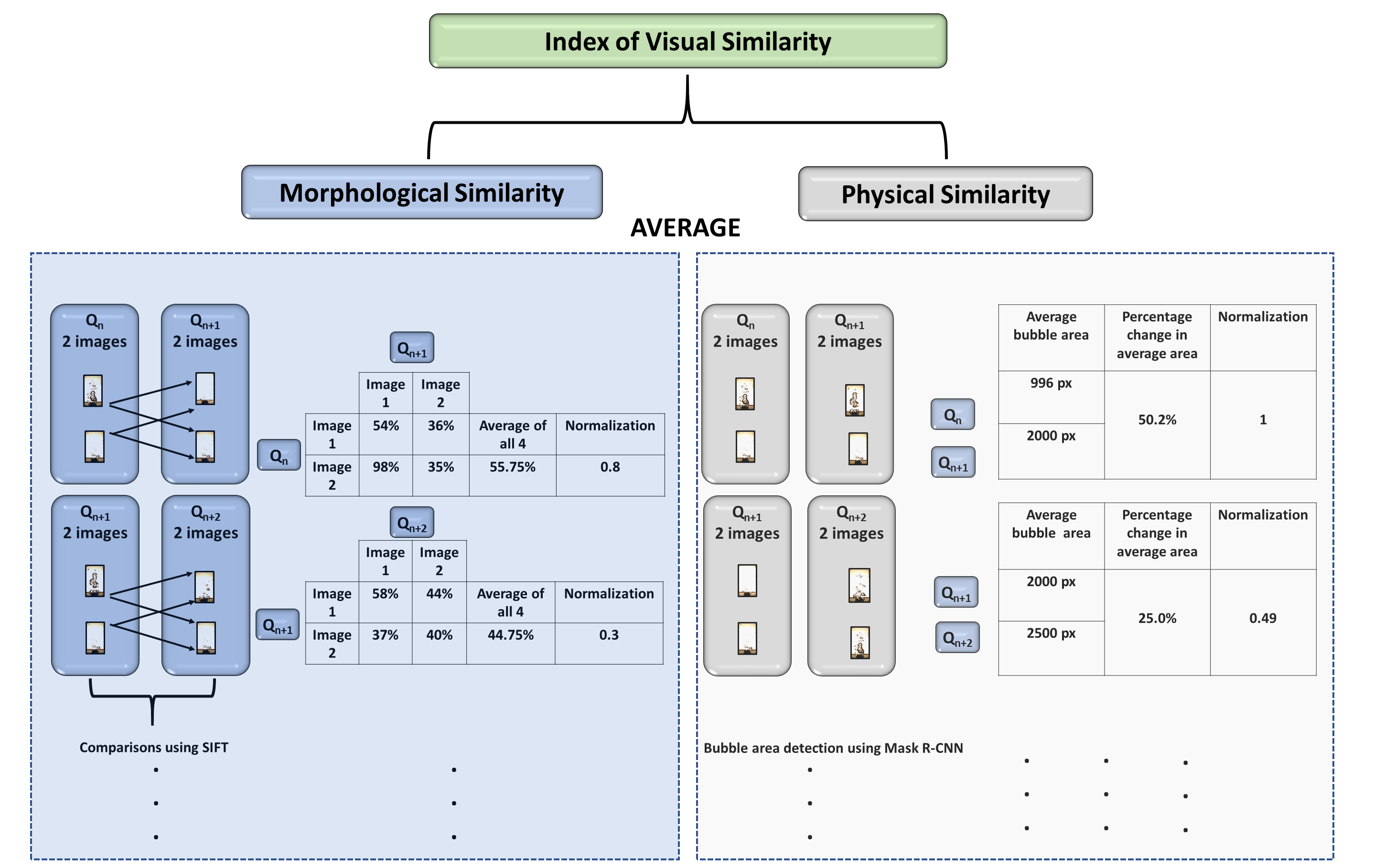}}
    \caption{Procedure for calculating the Index of Visual Similarity (IVS)}
    \label{fig11}
\end{figure}

\begin{figure}[htbp]
    \centering
    \begin{subfigure}[b]{0.45\textwidth}
        \centering
        \includegraphics[scale=0.52]{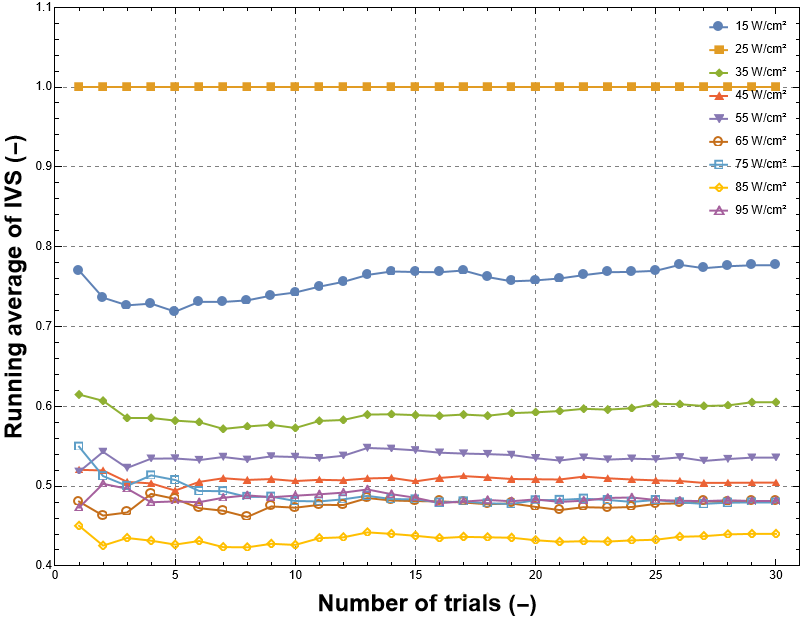}
        \caption{Selection of number of trials for dataset 1 }
        \label{fig:subfig11}
    \end{subfigure}
    \hfill
    \begin{subfigure}[b]{0.45\textwidth}
        \centering
        \raisebox{0pt}{\includegraphics[scale=0.32]{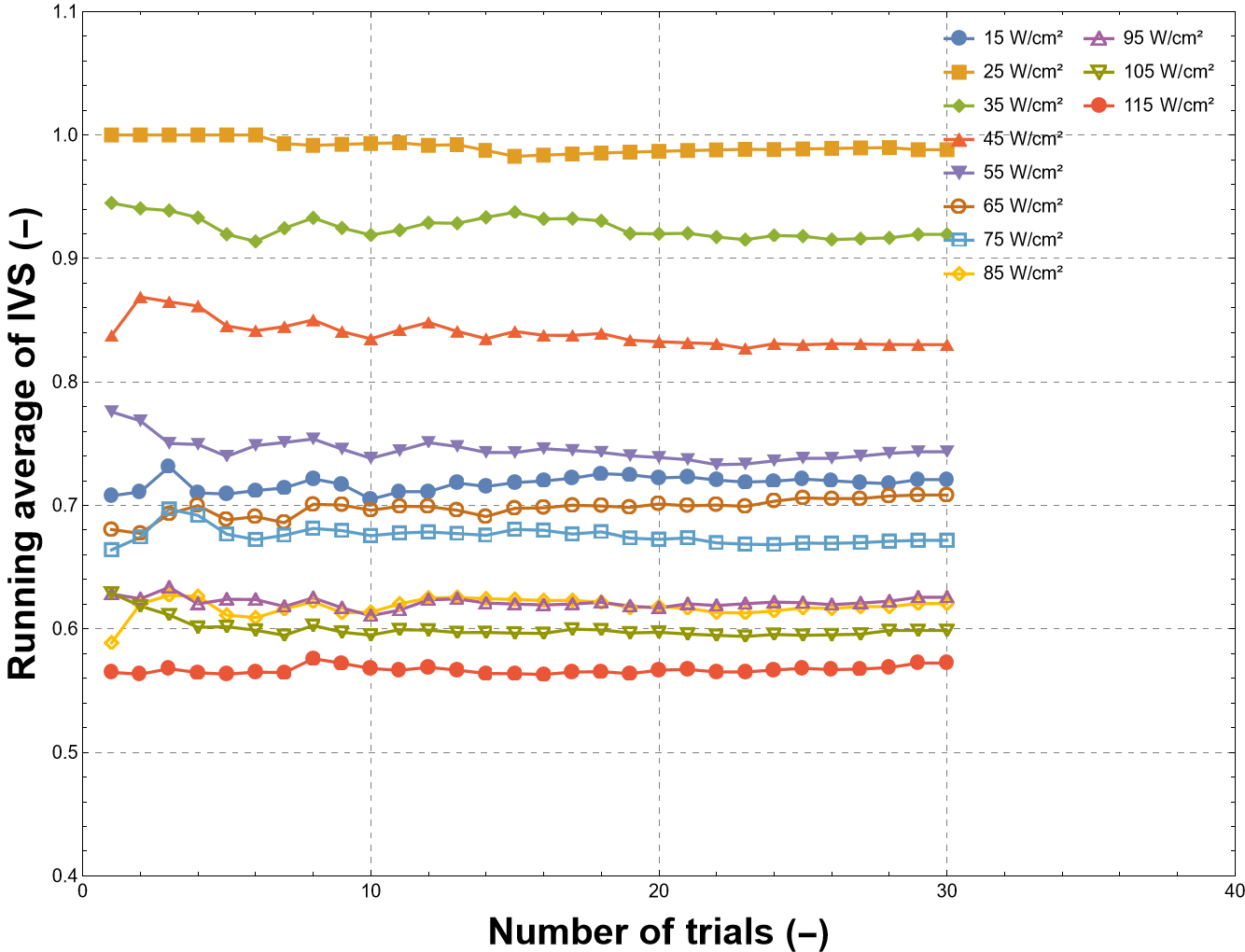}} % Moves the second subfigure up by 10pt
        \caption{Selection of number of trials for dataset 2}
        \label{fig:subfig22}
    \end{subfigure}
    \caption{Empirical selection of the number of trials needed to ensure convergence of IVS computation for (a) Dataset-1 and (b) Dataset-2.}
    \label{fig12}
\end{figure}

The choice of $\Delta Q$ is critical and should align with the desired resolution for distinguishing visual features. For each heat flux, two images are randomly selected from a pool of 3000 high-speed frames captured during steady-state conditions. These image pairs are used to compute both morphological and physical similarities, the average of which yields the IVS for that particular comparison. Selecting only two images per heat flux enables rapid computation and minimizes data requirements, allowing for faster processing. However, this minimalistic approach also introduces potential inaccuracies due to the visually transient nature of boiling. To enhance robustness, the entire process is repeated seven times with independently sampled image pairs. The final IVS is computed as the average across these seven trials. 
The number of repetitions was empirically determined by observing the convergence of the running average of IVS. As shown in fig. \ref{fig12}, the metric stabilizes beyond seven trials, justifying the selection.

Two distinct similarity metrics, each capturing a different class of visual features, are computed as follows:
\subsubsection {Morphological Similarity} 
Morphological similarity quantifies the degree of common, identifiable high-level features (such as bubble form and spatial proximity) between still frames corresponding to two different heat fluxes. To compute this between $n^{th}$ and $(n+1)^{th}$ heat fluxes, each of the four possible image pairings (two images from each heat flux) is analyzed to obtain a matching score. 

The similarity score for a given image pair is determined using a combination of the Scale-Invariant Feature Transform (SIFT) \cite{b32} and Brute-Force Matching.

Initially, all images are converted to grayscale, after which SIFT is used to extract keypoints and corresponding descriptors from each image. Feature matching is then performed using Brute-Force Matching with k-nearest neighbors (k = 2). The Lowe’s ratio test is applied with a threshold of 0.88 to eliminate false matches, by retaining only those matches for which the nearest neighbor is at least 12\text{\%} better than the second-closest match.

A similarity score for an image pair is calculated as the ratio of these “good matches” to the total number of keypoints in the first image, thereby reflecting the morphological and structural resemblance between the images. An average similarity score for the heat flux pair $Q_n$ and $Q_{n+1}$ is obtained by averaging the similarity scores across all four image combinations.
\begin{equation} \bar{M}_{(n,n+1)} = \sum_{t=1}^{4} M_t \end{equation}
where, $\bar{M}_{(n,n+1)}$= Averaged similarity score for comparison between $n^{th}$ and $n+1^{th}$ heat fluxes
and M= Similarity score for an image combination from $n^{th}$ and $n+1^{th}$ heat fluxes .
Final morphological similarity is calculated by normalizing the averaged similarity score as follows:
\begin{equation}(Morphological\text{ }similarity ) _{n}=\frac{\bar{M}_{n,n+1}*100}{Max (\bar{M}_{(1,2)},\bar{M}_{(2,3)}...\bar{M}_{m,m+1})} \end{equation}
where, m= total number of heat flux recorded.
\subsubsection{Physical Similarity}
Physical similarity aims to capture variations in visual features that correspond to physically meaningful quantities influenced by heat flux -in this case, the volume of vapor generated. Since direct volumetric estimation is challenging in two-dimensional imaging, the vapor volume is approximated by calculating the two-dimensional projected vapor area from still frames.

To accomplish this, we implemented an instance segmentation framework based on Mask R-CNN for precise identification and localization of bubble regions within each image \cite{b33}. A custom dataset comprising a combination of 35 high and low-resolution images sourced from diverse origins, including in-house experimental data, was manually annotated to delineate the exact bubble boundaries.

The Mask R-CNN model was fine-tuned on this annotated dataset, initialized with pre-trained weights from the MS COCO dataset. To enable domain adaptation, the final classification and mask prediction layers of the original model were replaced with layers specific to bubble detection. The model was trained over 25 epochs, optimizing both the bounding box and mask outputs to achieve accurate instance-level segmentation of bubbles.

\begin{figure}[htbp]
    \centerline{\includegraphics[scale=0.28]{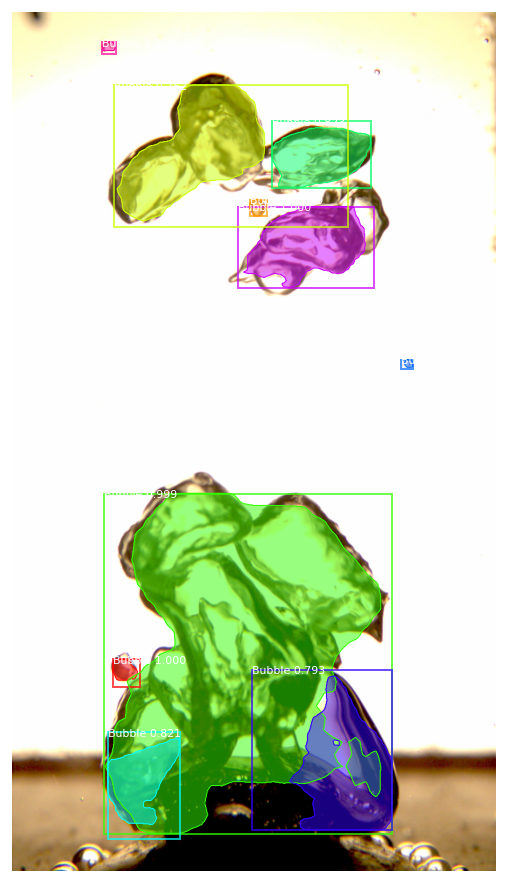}}
    \caption{Detected bubble instances in a high-speed image at 100 W/cm² using Mask R-CNN}
    \label{fig:subfig1}
\end{figure}

\begin{figure}[htbp]
    \centerline{\includegraphics[scale=0.30]{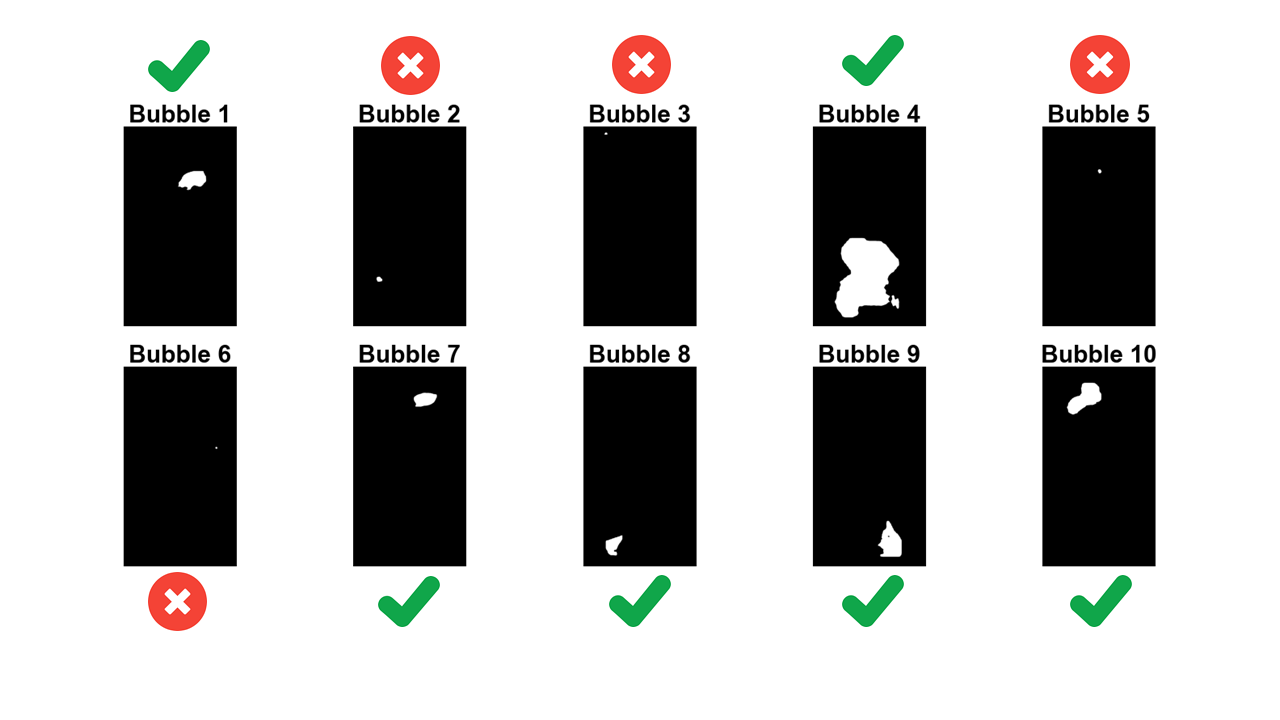}}
    \caption{Corresponding binary segmentation masks used for vapor area estimation. Only masks with an area greater than the average bubble size are counted further. This is done to exclude the smaller bubbles generated from guard heaters.}
    \label{fig:subfig2}
\end{figure}

The bubble area is computed from the instance segmentation masks generated by the trained Mask R-CNN model. Each segmented bubble is represented as a binary mask, where pixels belonging to the bubble are assigned a value of 1, and background pixels are set to 0. To quantify the area, we extract these binary masks corresponding to each detected bubble in an image. These masks share the same dimensions as the original input image.
The area of an individual bubble, $A_b$ is calculated by summing the number of pixels within its binary mask that have a value of 1. This is mathematically expressed as:

\begin{equation}
A_b = \sum_{i=1}^{H} \sum_{j=1}^{W} b(i, j)
\end{equation}

where $b(i, j)$represents the binary value (1 or 0) of the mask at pixel coordinates $(i,j)$, and $H$ and $W$ are the height and width of the image. The total number of non-zero pixels directly corresponds to the projected 2D area of the bubble in pixel units.

For any $n^{th}$ heat flux, two images are randomly selected from the available dataset, and the total vapor area is computed by summing the areas of the bubbles that are larger than the average bubble size in each image. This is done to exclude the smaller bubbles generated from guard heaters. 
\begin{equation}A_{\text{vapor}} = \sum_{k=1}^{m} A_k, \quad \text{where } A_k > \bar{A}\end{equation}
where,

  $A_k$: Area of the $k^{\text{th}}$ bubble, where $A_k > \bar{A}$ (i.e., larger than the average bubble area $\bar{A}$)
  $m$: Total number of bubbles satisfying the condition $A_k > \bar{A}$

The total vapor area is then averaged for these two randomly picked images, representing the vapor volume corresponding to that heat flux as shown below: 
\begin{equation}\bar{A}_{\text{n,vapor}} = \frac{1}{2} \left( A_{\text{n,vapor}}^{(image\text{}1)} + A_{\text{n,vapor}}^{(image\text{}2)} \right)\end{equation}
The same process is repeated for the $(n+1)^{th}$ heat flux for comparison.
 
The physical similarity between $n^{th}$ and $(n+1)^{th}$ heat flux is quantified by evaluating the percentage change in average total vapor area representing these two heat fluxes. The inverse of this percentage change is defined as the physical similarity, such that a smaller difference in vapor area yields a higher similarity score.
\begin{equation}Physical\text{ }Similarity =100- {\frac{|\bar{A}_{\text{n,vapor}}-\bar{A}_{\text{n+1,vapor}}|*100} {\bar{A}_{\text{n,vapor}}}}\end{equation}

To mitigate the influence of transient fluctuations and enhance robustness, the entire process is repeated over seven independent trials, each time selecting a new pair of random images. The final physical similarity score is computed as the mean across these trials, ensuring statistical reliability.

Finally IVS is calculated as:
\begin{equation}IVS= 0.5*(Physical\text{ }Similarity)\text{ }+\text{ }0.5*(Morphological\text{ }Similarity)\end{equation}

\subsection{\label{sec:level2}Measurement uncertainty analysis}

Major source of uncertainty in the measurement of heat flux is the uncertainty associated with the temperature measurement. Root sum of squares (RSS) \cite{b36} method was used to quantify the uncertainty propagation from the thermocouple's uncertainty.  
Measurement of temperature was carried out using T-type thermocouples and uncertainty associated with the used thermocouples was $\pm 0.5$ °C and the tolerance limit of machining is $\pm $ 0.25 mm. Hence the maximum uncertainty in measured heat flux is  7.68 W/cm$^2$ 
%as can be seen in fig. \ref{figad}.

The uncertainty in the thermocouples used to measure the temperature in the experiment was $\pm 0.5$ °C. 
Uncertainty in the heat flux measurement is calculated as :

\begin{equation}
    u_{\Delta T^*} = \sqrt{(3u_T)^2 + (4u_T)^2 + (u_T)^2}
\end{equation}
\begin{equation}
    \frac{u_{q}}{q} = \sqrt{\left(\frac{u_{\Delta T^*}}{\Delta T^*}\right)^2 + \left(\frac{u_{\Delta x}}{\Delta x}\right)^2} 
\end{equation}

The maximum uncertainty in the heat flux and wall superheat were found to be $\pm$ 7.68 W/cm$^2$ and $\pm$ 1.02°C respectively.  
%\begin{figure}[htbp]
 %   \centerline{\includegraphics[scale=0.50]{Uncertainty in measurement.png}}
  %  \caption{Quantitative analysis of uncertainty in heat flux measurements}
   % \label{figad}
%\end{figure}

\subsection{Metric to asses change in heat transfer coefficient and uncertainty associated}\label{phi}

% Discuss in detail the significance of IVS with the evolution in time and the blurred image gap and also stress upon the fact that the explored gap is 20W/cm2 and around that range and needs to be scrutinized by a formulated paradigm
HTC serves as a key performance parameter for indirectly assessing the relative contributions of various heat transfer mechanisms. Each mechanism (ranging from single-phase vapor transport, which contributes the least, to contact line evaporation, which is the most efficient) has a distinct capacity to transfer heat. Therefore, variations in HTC can reflect shifts in the dominant mechanisms at play. To capture these dynamics, the HTC is computed at each heat flux as follows:
\begin{equation}
h_i = \frac{q_i}{T_{s,i}}
\end{equation}
where \( q_i \) is the heat flux and \( T_{s,i} \) is the surface superheat temperature at the \( i \)-th measurement.

\begin{figure}[htbp]
    \centerline{\includegraphics[scale=0.5]{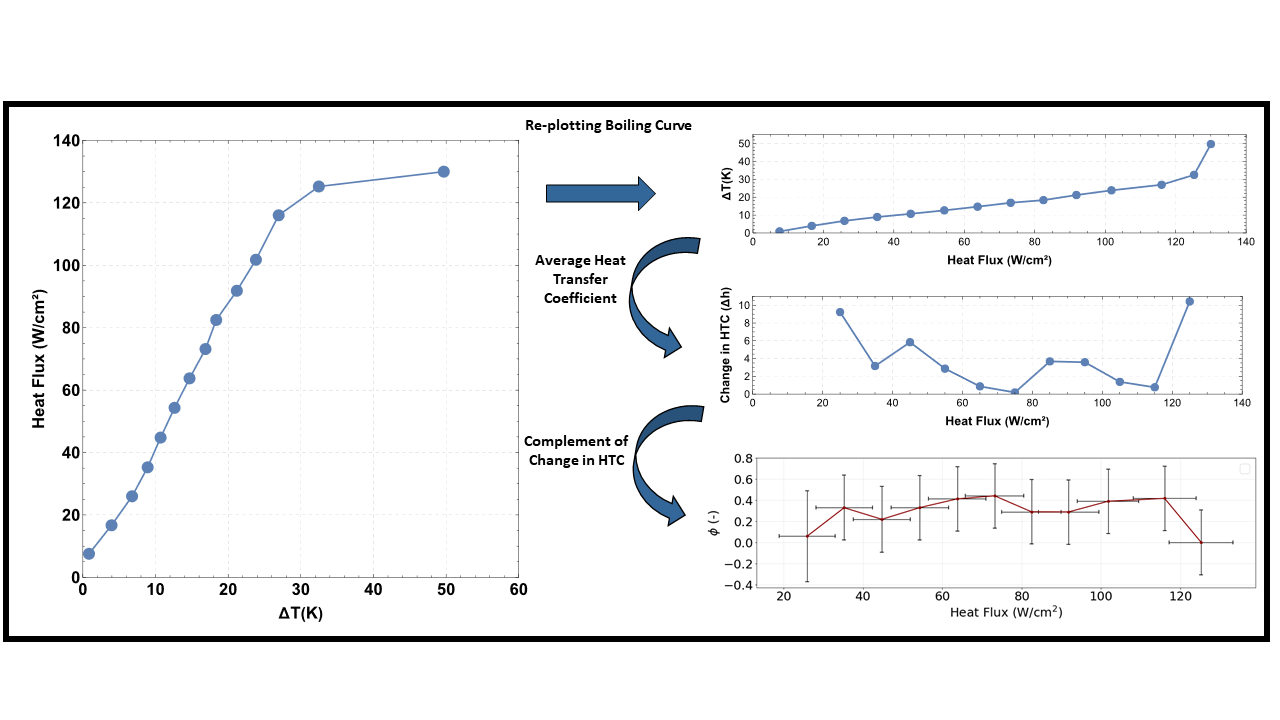}}
    \caption{Metric assessing relative changes in heat transfer coefficient (HTC) for Dataset-2}
    \label{fig14}
\end{figure}

The relative uncertainty in \( h_i \) is given by standard uncertainty propagation:
\begin{equation}
\left( \frac{\delta h_i}{h_i} \right)^2 = \left( \frac{\delta q_i}{q_i} \right)^2 + \left( \frac{\delta T_{s,i}}{T_{s,i}} \right)^2
\end{equation}
\begin{equation}
\Rightarrow \delta h_i = h_i \cdot \sqrt{ \left( \frac{\delta q_i}{q_i} \right)^2 + \left( \frac{\delta T_{s,i}}{T_{s,i}} \right)^2 }
\end{equation}

This relative change is then complemented with respect to its maximum value, producing a metric that aligns with the interpretation of IVS (represented as $\phi$ : higher values indicate greater similarity i.e., less fluctuation in HTC), while lower values indicate more pronounced changes. This transformation enables a direct and intuitive comparison between IVS and the thermal behavior of the system.

Change in h:
\begin{equation}
\Delta h_i = h_i-h_{i-1}
\end{equation}

{Uncertainty in \texorpdfstring{$\Delta h_i$}{Delta hi}
Using error propagation:
\begin{equation}
\delta(\Delta h_i) /h_i = \sqrt{
\left( \frac{ \delta h_i }{ h_i} \right)^2 +
\left( \frac{ \delta h_{i-1}}{ h_{i-1}} \right)^2
}
\end{equation}

The metric \( \phi \) is defined as follows:
\begin{equation}
\phi_i = \max(\Delta h_j) - \Delta h_i \quad \text{for all } j
\end{equation}

Uncertainty in \texorpdfstring{$\phi$}{phi}
Let \( \Delta h_{\max} = \Delta h_k \), then:
\begin{equation}
\delta \phi_i = \sqrt{ \left( \delta(\Delta h_k) \right)^2 + \left( \delta(\Delta h_i) \right)^2 }
\end{equation}

Respective plots for dataset-2 are illustrated in fig. \ref{fig14}.

\section{Summary and conclusions}

This work presents a novel, data-driven framework to quantify visual changes in boiling phenomena using the IVS. This hybrid metric captures both structural and physically relevant features from high-speed image data. IVS combines two independently calculated sets of features: morphological similarity, which captures changes in the bubble structure and spatial distribution, and physical similarity, which estimates changes in vapor volume through deep learning-based instance segmentation.

The methodology was validated across two datasets obtained using different surfaces and visualization techniques to ensure generalizability. A key strength of IVS lies in its ability to detect subtle changes in boiling patterns that align strongly with known transitions in boiling regimes, such as the onset of nucleate boiling and proximity to CHF. This has been reinforced by comparing the relative change in the HTC and surface superheat with IVS, confirming its physical significance and potential as a non-intrusive diagnostic tool.

IVS offers a scalable general solution without rigorous model training that requires only minimal image data, enabling rapid and robust assessment of the boiling regime. While the current study focuses on regime identification, future extensions aim to refine the use of IVS further to estimate each heat flux precisely, thereby opening new avenues for real-time thermal system monitoring and control.

\end{document}